\documentclass[twocolumn,showpacs,preprintnumbers,prl,floatfix]{revtex4}
\usepackage{amsmath}
\usepackage{amssymb}
\usepackage{graphicx}
\usepackage{dcolumn}
\usepackage{epsfig,color}
\usepackage{bm}
\usepackage{longtable}
\usepackage{rotating}

\newcommand{\ba}{\begin{eqnarray}}
\newcommand{\ea}{\end{eqnarray}}
\begin{document}

\title{Evidence for Triangular $D'_{3h}$ Symmetry in $^{13}$C}
\author{R. Bijker} 
\affiliation{Instituto de Ciencias Nucleares, Universidad Nacional Aut\'onoma de M\'exico, 
AP 70-543, 04510 M\'exico DF, M\'exico}
\author{F. Iachello}
\affiliation{Center for Theoretical Physics, Sloane Laboratory, 
Yale University, New Haven, Connecticut 06520-8120, USA}

\begin{abstract}
We derive the rotation-vibration spectrum of a $3\alpha+1$ neutron(proton)
configuration with triangular $D_{3h}$ symmetry by exploiting the properties
of the double group $D_{3h}^{\prime }$, and show evidence for this symmetry
to occur in the rotation-vibration spectra of $^{13}$C. 
Our results, based on purely symmetry considerations, provide benchmarks for 
microscopic calculations of the cluster structure of light nuclei.
\end{abstract}

\maketitle

The cluster structure of light nuclei is a long standing problem which goes
back to the early times of nuclear physics \cite{wheeler}. Recently, there
has been renewed interest in this problem due to new measurements in $^{12}$%
C \cite{itoh,freer,zimmermann,marin-lambarri}, showing
evidence for the occurrence of $D_{3h}$ triangular symmetry in this nucleus.
Most applications of cluster models has been so far limited to $k\alpha $
nuclei, that is nuclei composed of $k$ $\alpha $-particles, with $k=2$, 3, 4,
which display $Z_{2}$ ($^{8}$Be), $D_{3h}$ ($^{12}$C) and $T_{d}$ ($^{16}$O)
symmetry \cite{brink-1,brink-2}. Study of stuctures composed of $k$ $%
\alpha $-particles plus $x$ additional nucleons, simply denoted here by $%
k\alpha +x$ nuclei, has been hindered by the lack of understanding of the
single-particle motion in an external field with arbitrary discrete
symmetry, $G$, and, especially, by the lack of explicit construction of
representations of the double group, $G^{\prime}$, which allows the
enlargement of tensor (bosonic) representations of the group $G$ to cases in
which there is one fermion, the so-called spinor (fermionic) representations. 
Recently we have started a systematic investigation of both problems. 
The study of the splitting of single-particle levels in an external field with 
$Z_{2}$, $D_{3h}$ and $T_{d}$ symmetry was carried out in Ref.~\cite{dellarocca-1}. 
The construction of representations of the double group $Z_{2}^{\prime}$ is 
trivial since the 2$\alpha $ structure possessing this symmetry is a dumbbell 
configuration with axial symmetry \cite{dellarocca-2}. The construction of 
representations of the double groups $D_{3h}^{\prime}$ and $T_{d}^{\prime}$ 
is more complicated. Although done for applications to crystal physics by Koster 
\textit{et al.} \cite{koster} and molecular physics by Herzberg \cite{herzberg}, 
to the best of our knowledge, it has never been done for applications 
to nuclear physics. In this article, we report the results of our investigation 
of the double group $D_{3h}^{\prime}$ and, in an application to the nucleus $^{13}$C, 
we present evidence for the occurrence of $D_{3h}^{\prime}$ symmetry in nuclear physics. 

The double group $D_{3h}^{\prime}$ has three spinor representations, denoted by 
Koster as $\Gamma_{7}$, $\Gamma_{8}$, $\Gamma_{9}$ \cite{koster} and by Herzberg 
as $E_{1/2}$, $E_{5/2}$, $E_{3/2}$ \cite{herzberg}. We prefer, for applications 
to nuclear physics, to denote the three representations by 
$E_{1/2}^{(+)} \equiv \Gamma_{7} \equiv E_{1/2}$, 
$E_{1/2}^{(-)} \equiv \Gamma_{8}\equiv E_{5/2}$, 
$E_{3/2} \equiv \Gamma_{9} \equiv E_{3/2}$, and label the states by 
$\left\vert \Omega,K,J \right\rangle$, where $\Omega$ labels the
representations of $D_{3h}^{\prime}$, and $K$ and $J$ are half integers representing 
the projection $K$ of the total angular momentum $J$ on a body-fixed axis.  
The allowed values of $K^P$ for each one of the spinor representations are given by 
\begin{equation}
\begin{array}{lllll}
\Omega = E_{1/2}^{(+)}: &\hspace{0.5cm}& K^P=1/2^+ \quad \mbox{and}& \\
&& K=3n \pm \frac{1}{2} && P=(-)^{n}   \\ 
\Omega = E_{1/2}^{(-)}: && K^P=1/2^- \quad \mbox{and}& \\
&& K=3n \pm \frac{1}{2} && P=(-)^{n+1} \\
\Omega = E_{3/2}: && K^P=(3n - \frac{3}{2})^{\pm} && 
\end{array}
\label{angmom}
\end{equation}
with $n=1,2,3,\ldots,$ and $K > 0$. The angular momenta of each $K$ band are 
given by $J=K,K+1,K+2,\ldots$. Note the double degeneracy $K^P=K^{\pm}$ for  
the representation $E_{3/2}$ (parity doubling).

This classification allows one to construct the rotational spectrum 
of a triangular configuration of three $\alpha$ particles dragging along an additional 
proton or neutron. The rotational formula is
\begin{equation}
E_{rot}(\Omega,K,J) = \varepsilon_{\Omega} + A_{\Omega} 
\left[ J(J+1) + b_{\Omega} K^{2} + a_{\Omega} g_{\Omega }(J) \right] ~,
\label{erot}
\end{equation}
where $\varepsilon_{\Omega}$ is the intrinsic energy \cite{dellarocca-1}, $A_{\Omega} = \hbar^{2}/2\Im$ 
is the inertial parameter, $b_{\Omega}$ is a Coriolis term, and $a_{\Omega}$ is the so-called decoupling 
parameter with $g_{\Omega }(J)= \delta_{K,1/2} (-1)^{J+1/2}(J+1/2)$. The latter term applies only to 
representations $\Omega \equiv E_{1/2}^{(\pm)}$ and $K^P=1/2^{\pm}$.

\begin{figure}
\centering
\includegraphics[width=\linewidth]{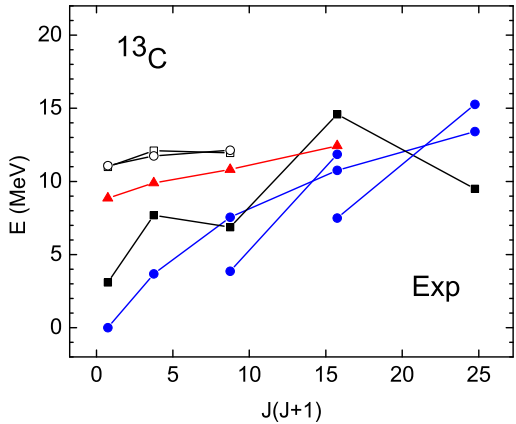}
\caption{The rotational spectra of $^{13}$C. 
Energy levels \cite{ajzenberg-selove} are plotted as a function of $J(J+1)$. 
For states below 10 MeV, our assignment of rotational bands is unambiguous. 
For states above 10 MeV, our assignment is tentative. 
\label{C13exp}}
\end{figure}

\begin{figure}
\centering
\includegraphics[width=\linewidth]{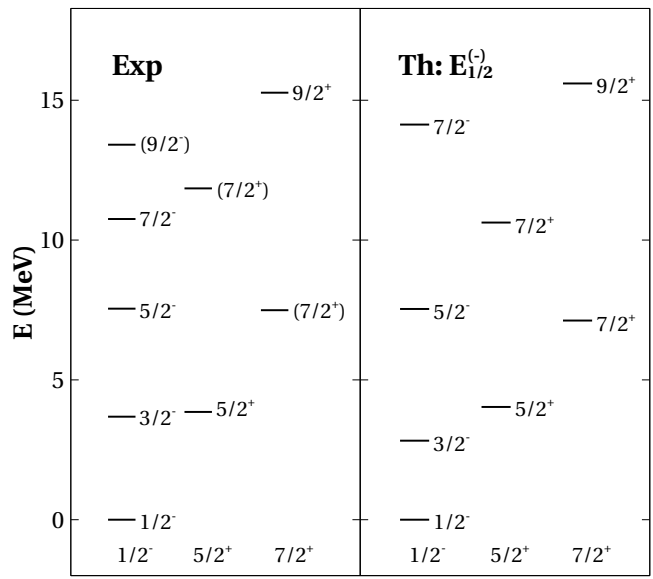}
\caption{Comparison between experimental and theoretical energies 
for the ground state band assigned to the representation $\Omega=E_{1/2}^{(-)}$ of 
$D_{3h}^{\prime}$. The values of $K^P$ are given at the bottom of the figure. 
The energies are calculated using Eq.~(\ref{erot}) with
$A_{\Omega}=0.942$ MeV, $b_{\Omega}=-0.62$ and $a_{\Omega}=0$.} 
\label{eminus}
\end{figure}

The rotational spectra of $^{13}$C are shown in Fig.~\ref{C13exp} 
where the experimental levels are plotted as a function of $J(J+1)$.
The ground state band has $K^{P}=1/2^{-}$ and it can be assigned to the
representation $\Omega=E_{1/2}^{(-)}$ of $D_{3h}^{\prime }$ 
(blue lines and filled circles). As seen from Eq.~(\ref{angmom}),
this representation contains also $K^{P}=5/2^{+}$ and $7/2^{+}$ bands.
Both of them appear to be observed as shown in Fig.~\ref{eminus}.
The observation of low-lying positive parity states with $K^{P}=5/2^{+}$ and 
$7/2^{+}$ is crucial evidence for the occurrence of $D_{3h}^{\prime }$
symmetry. In the shell model, positive parity states are expected to occur
at much higher energies since they come from the $s$-$d$ shell. They were not 
considered in the original calculation of Cohen and Kurath \cite{kurath}. 
In more recent calculations which include $(0s)^3(1p)^{10}$ plus $(0s)^4(1p)^8(2sd)^1$ 
configurations they are brought down by lowering the energy of the $2s_{1/2}$ level 
from 11 MeV to 5.43 MeV \cite{millener1} or 5.52 MeV \cite{lee}, and by adjusting the 
p-h interaction \cite{millener1}. 

\begin{figure}
\centering
\includegraphics[width=\linewidth]{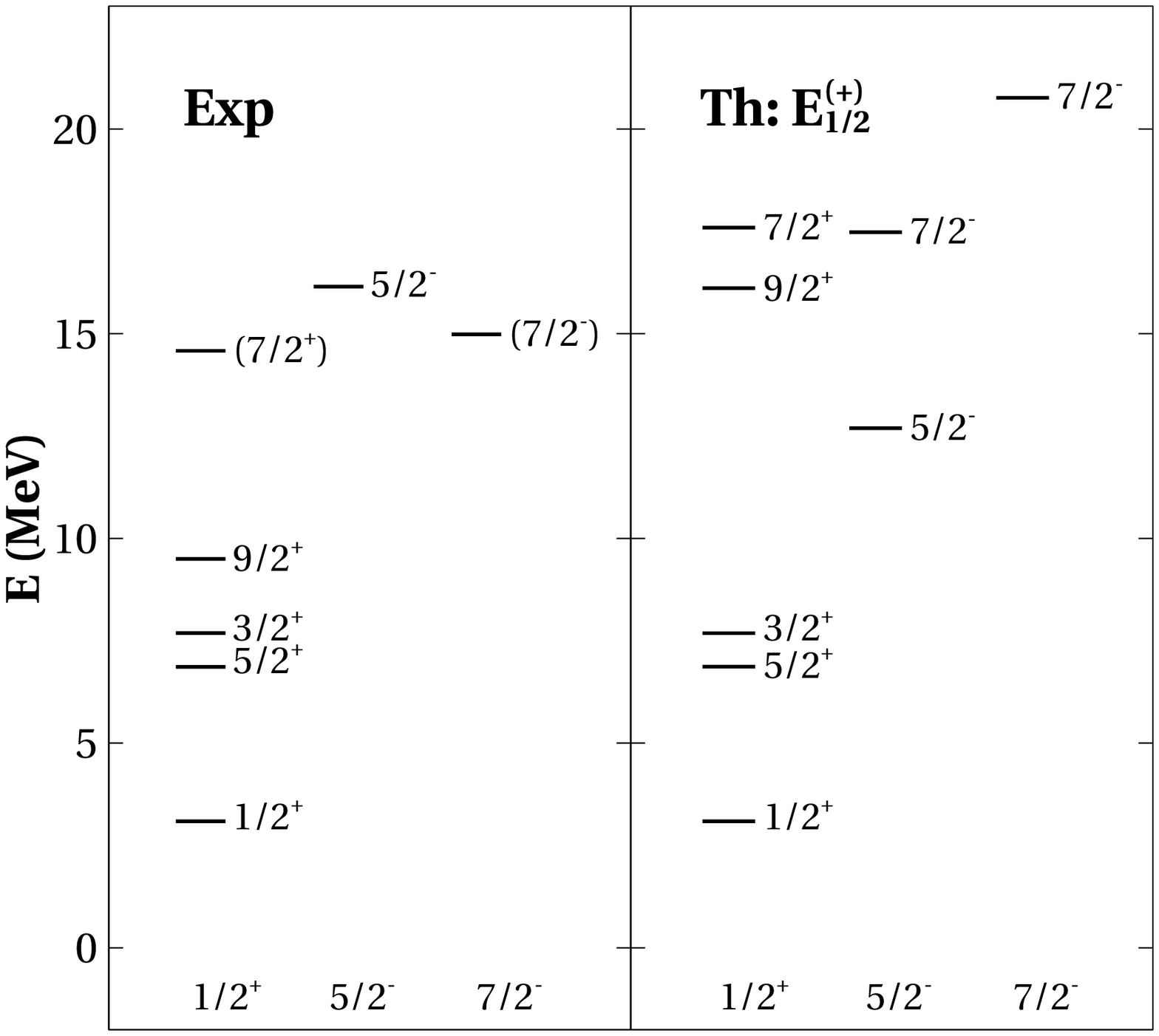}
\caption{As Fig.~\ref{eminus}, but for the first excited band assigned to the representation 
$\Omega=E_{1/2}^{(+)}$ of $D_{3h}^{\prime}$. The energies are calculated using 
Eq.~(\ref{erot}) with $A_{\Omega}=0.684$ MeV, $b_{\Omega}=0.80$, $a_{\Omega}=1.24$ 
and $\varepsilon=3.848$ MeV.} 
\label{eplus}
\end{figure}

The first excited rotational band has $K^{P}=1/2^{+}$. It can be assigned to
the representation $\Omega=E_{1/2}^{(+)}$ of $D_{3h}^{\prime}$ (black line and filled 
squares). This band has a large decoupling parameter, $a_{\Omega}=1.24$. 
According to Eq.~(\ref{angmom}),  
this representation contains also $K^{P}=5/2^{-}$ and $K^{P}=7/2^{-}$ bands.
The evidence for these bands is meager, as they are expected to lie at high
energy. There is some tentative evidence for the $K^{P}=5/2^{-}$ band at
energies $>15$ MeV, but no evidence for the $K^{P}=7/2^{-}$. This appears to
indicate that the Coriolis coefficient $b_{\Omega}$ is less negative than
that of the $E_{1/2}^{(-)}$ band (or even positive). Assuming a value of 
$b_{1/2^{+}}=0.80$ we calculate the $K^{P}=5/2^{-}$ bandhead at $\sim 13$ MeV 
and the $K^{P}=7/2^{-}$ bandhead at $\sim 20$ MeV. 
This situation is shown in Fig.~\ref{eplus}. 

The experimental value of the energy difference $E(1/2_1^+)-E(1/2_1^-)=3.089$ MeV 
is further evidence of $D_{3h}^{\prime}$ symmetry in $^{13}$C. From Fig.~11 of 
Ref.~\cite{dellarocca-1} we can estimate this value to be $\sim 2.0$ MeV. 
Again, in the shell model the $1/2^+$ state comes from the $s$-$d$ shell, and is brought 
down by the lowering of the $2s_{1/2}$ level as mentioned in the paragraph above.

The expected vibrational spectra can be obtained by coupling the representations 
of the fundamental vibrations of the triangular configuration with symmetry 
$A_{1}^{\prime}$ and $E^{\prime}$ \cite{bijker-2} to the intrinsic states with 
$E_{1/2}^{(-)}$ and $E_{1/2}^{(+)}$ symmetry. From the multiplication table of 
$D_{3h}^{\prime}$ one obtains \cite{koster,herzberg} 
\begin{eqnarray}
A_{1}^{\prime} \otimes E_{1/2}^{(\pm)} &=& E_{1/2}^{(\pm)} ~,  
\nonumber\\
E^{\prime} \otimes E_{1/2}^{(\pm)} &=& E_{3/2} \oplus E_{1/2}^{(\mp)} ~. 
\label{vibrations}
\end{eqnarray}
For each intrinsic state, one expects three states, $\Omega=E_{1/2}^{(-)}$, $E_{3/2}$, 
$E_{1/2}^{(+)}$ for the intrinsic state with $E_{1/2}^{(-)}$ symmetry, and 
$\Omega=E_{1/2}^{(+)}$, $E_{3/2}$, $E_{1/2}^{(-)}$ for $E_{1/2}^{(+)}$. 
We denote the corresponding vibrational quantum numbers  
by $v_{1\Omega}$, $v_{2\Omega}$, $v_{3\Omega}$, respectively, where, for simplicity 
of notation, we have omitted the label of the vibronic angular momentum $l$. 
In the analysis of the vibrational states, it is convenient to remove the zero-point
energy. The vibrational formula, to lowest order in the vibrational quantum
numbers (harmonic limit) is 
\begin{equation}
E_{vib}(\Omega;v_{1\Omega},v_{2\Omega},v_{3\Omega})=\omega_{1\Omega} v_{1\Omega} 
+ \omega_{2\Omega} v_{2\Omega} + \omega_{3\Omega} v_{3\Omega} ~.
\end{equation}

The vibration $A_{1}^{\prime }$ in $^{12}$C plays an important role in nuclear astrophysics 
since it is associated with the so-called Hoyle state. According to Eq.~(\ref{vibrations}) 
we expect Hoyle states also in $^{13}$C. Indeed, the Hoyle band built on top of the ground 
state $E_{1/2}^{(-)}$ representation appears to have been observed in $^{13}$C starting at 
an energy of 8.860 MeV (red line and filled triangles in Fig.~\ref{C13exp}), 
which is slightly higher than that of the Hoyle state in $^{12}$C (7.654 MeV). 
The moment of inertia of this band is similar to that of
the Hoyle band in $^{12}$C, which is further evidence for the occurrence of 
$D_{3h}^{\prime }$ symmetry in $^{13}$C. In Fig.~\ref{C13exp}, one can also observe two
additional bands with $K^{P}=1/2^{+}$ and $K^{P}=1/2^{-}$ starting at 10.996
MeV and 11.080 MeV. Because many states with these values of $K^{P}$ are
expected in this region, no firm assignments can be made, but it is very
likely that these bands are the vibrations $E_{1/2}^{(+)}$ and $E_{1/2}^{(-)}$ 
of Eq.~(\ref{vibrations}).

In the region $E \sim 10$ MeV, one expects additional rotational bands. 
Evidence for two rotational bands with $K^P=3/2^{\pm}$ has been reported \cite{Freer84}, 
starting at 9.90 MeV ($3/2^-$) and 11.08 MeV ($3/2^+$), respectively. These bands can 
be assigned to the representation $\Omega=E_{3/2}$ of $D_{3h}^{\prime}$ (see Eq.~(\ref{angmom})), 
and split into its two components by Coriolis and other interactions. These bands were suggested 
to arise from $^{9}\mbox{Be}+\alpha$ configurations \cite{Milin}. A discussion of these bands 
will be presented in a longer publication. 

\begin{figure}
\centering
\includegraphics[width=\linewidth]{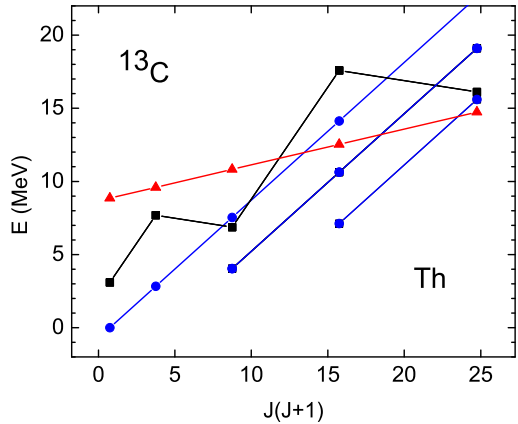}
\caption{Rotational spectra in $^{13}$C expected on the basis of 
$D_{3h}^{\prime}$ symmetry.}.
\label{C13th}
\end{figure}

The situation for rotational and vibrational bands in $^{13}$C is summarized in Fig.~\ref{C13th}. 
A comparison with the experimental spectrum in Fig.~\ref{C13exp} shows evidence for $D_{3h}^{\prime}$ 
symmetry in $^{13}$C. 

Further evidence for the occurrence of $D_{3h}^{\prime }$ symmetry in $^{13}$C 
is provided by electromagnetic transition rates and form factors in electron scattering. 
A complete analysis of electromagnetic transition rates and electromagnetic form factors in
electron scattering requires an elaborate calculation similar to that done
for $^{9}$Be and $^{9}$B in Ref.~\cite{dellarocca-2}. Here we limit
ourselves to the most important points. 

\begin{widetext}
$B(E\lambda)$ values in $k\alpha+x$ nuclei can be calculated using using Eq.~(25) 
of \cite{dellarocca-2} as
\ba
B(E\lambda;\Omega',J',K' \rightarrow \Omega,J,K) 
&=& \Big| \left<J',K',\lambda,K-K' | J,K \right> (\delta_{v,v'} G_{\lambda}(\Omega,\Omega') 
+ \delta_{\Omega,\Omega'} G_{\lambda,c}) 
\nonumber\\
&& + (-)^{J+K} \left<J',K',\lambda,-K-K' | J,-K \right> (\delta_{v,v'} \tilde{G}_{\lambda}(\Omega,-\Omega') 
+ \delta_{\Omega,-\Omega'} G_{\lambda,c}) \Big|^2 ~.
\ea
\end{widetext}
Here $G_{\lambda}(\Omega,\Omega')$ represents the contribution of th single particle 
and $G_{\lambda,c}$ the contribution of the cluster. In $^{13}$C the single particle 
is a neutron and thus it does not contribute to electric transitions, except for $E1$ 
transitions affected by the center-of-mass correction as discussed in Eq.~(32) of 
\cite{dellarocca-2}. The cluster contribution is given by the $D_{3h}$ symmetry as \cite{bijker-2}
\ba
G_{\lambda,c} = Z \beta^{\lambda} \sqrt{\frac{2\lambda+1}{4\pi}} \, c_{\lambda} ~,
\ea
where the coefficients $c_{\lambda}$ are given by
$c_0=1$, $c_2=1/2$, $c_3=\sqrt{5/8}$ and $c_4=3/8$.
The value of $\beta$ extracted from the minimum in the elastic form factor of $^{12}$C is 
$\beta=1.74$ fm. With this value we calculate the $B(E\lambda)$ values given in 
Table~\ref{BEL}, where they are compared with experiment. 
Both experimental and theoretical values in $^{12}$C show that both states, 
$2_{1}^{+}$ and $3_{1}^{-}$, belong to the same rotational band of the triangle \cite{bijker-2},
representation $A_{1}^{\prime }$ of $D_{3h}$. Note in particular the large 
$B(E3)$ value that cannot be obtained in shell-model calculations without the introduction 
of large effective charges.
Similarly, the values in $^{13}$C show that the states $3/2_{1}^{-}$, $5/2_{1}^{-}$ and 
$5/2_{1}^{+}$ belong to the same rotational band with $\Omega=E_{1/2}^{(-)}$.  
Note also here the large $B(E3;5/2_{1}^{+} \rightarrow 1/2_{1}^{-})$ value.
This value is obtained in the cluster calculation without the use of effective charges.

\begin{table}
\centering
\caption{$B(EL)$ values in $^{12}$C and $^{13}$C in W.U. \cite{ajzenberg-selove}.}
\label{BEL}
\begin{tabular}{lccc}
\hline
\noalign{\smallskip}
& $B(EL)$ & Exp & Th \\ 
\noalign{\smallskip}
\hline
\noalign{\smallskip}
$^{12}$C & $B(E2;2_{1}^{+}\rightarrow 0_{1}^{+})$ & $4.65 \pm 0.26$ & $4.8$ \\ 
& $B(E3;3_{1}^{-}\rightarrow 0_{1}^{+})$ & $12 \pm 2$ & $7.6$ \\ 
\noalign{\smallskip}
\hline
\noalign{\smallskip}
$^{13}$C & $B(E2;3/2_{1}^{-}\rightarrow 1/2_{1}^{-})$ & $3.5 \pm 0.8$ & $4.8$ \\ 
& $B(E2;5/2_{1}^{-}\rightarrow 1/2_{1}^{-})$ & $3.1 \pm 0.2$ & $3.2$ \\ 
& $B(E3;5/2_{1}^{+}\rightarrow 1/2_{1}^{-})$ & $10 \pm 4$ & $4.3$ \\
\noalign{\smallskip}
\hline
\end{tabular}
\end{table}

\begin{figure}
\centering
\includegraphics[width=\linewidth]{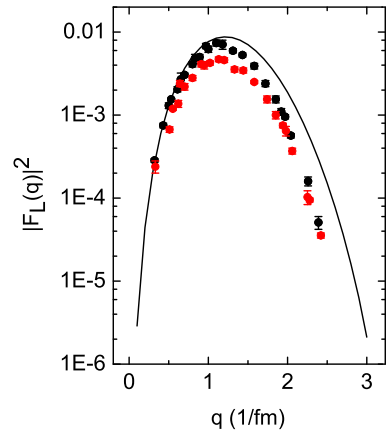}
\caption{Comparison between calculated and experimental \cite{millener2} longitudinal 
$E2$ form factors for the ground-state band of $^{13}$C, $1/2_1^- \rightarrow 5/2_1^-$ 
(black) and $1/2_1^- \rightarrow 3/2_1^-$ (red).}
\label{fffig}
\end{figure}

In the same way, form factors in electron scattering can be split into a single-particle 
and collective cluster contribution, $F(q)=F^{s.p.}(q)+F^{c}(q)$, as discussed in 
Sect.~3.6 of \cite{dellarocca-2}. For odd-neutron nuclei, the single particle does 
not contribute appreciably, except for multipolarity $E1$. The cluster contribution to the 
longitudinal electric form factors can be written as in Eq.~(46) of \cite{dellarocca-2} 
\ba
F_{\lambda}^c(q;J,K \rightarrow J',K')=\delta_{K,K'}Z \sqrt{\frac{2\lambda+1}{4\pi}} \, c_{\lambda}
\nonumber\\
\left< J,K,\lambda,0 \mid J',K' \right> j_{\lambda}(q\beta) \mbox{e}^{-q^2/4\alpha} ~,
\label{ff}
\ea
where $\alpha=0.56$ fm$^{-2}$ is obtained from electron scattering in $^{4}$He 
and $\beta=1.74$ fm from electron scattering in $^{12}$C. Using Eq.~(\ref{ff}), one 
can calculate all longitudinal form factors in a parameter independent way. 
The longitudinal form factors of the states $5/2_1^-$ and $3/2_1^-$ of the 
ground-state rotational bands are shown in Fig.~\ref{fffig} where they are compared 
with experimental data \cite{millener2}.
An important consequence of the cluster model is that the two form factors 
$1/2_1^- \rightarrow 3/2_1^-$ and $1/2_1^- \rightarrow 5/2_1^-$ have identical shapes,  
and identical $B(E2;\uparrow)$ values: 9.6 W.U. This is to a very good 
approximation seen in Fig.~\ref{fffig}.
The discrepancy at large momentum transfer is due to the fact that the value of $\beta$ 
appropriate to $^{12}$C has been used to make the calculation parameter free. A small 
renormalization of this value to $\beta=1.82$ fm reproduces the data perfectly.

\begin{figure}
\centering
\includegraphics[width=0.5\linewidth]{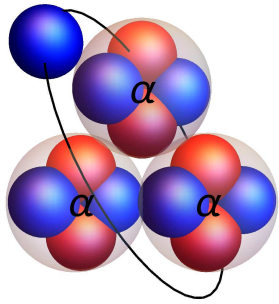}
\caption{Molecular-like picture of $^{13}$C.}
\label{3alphan}
\end{figure}

In conclusion, both the rotation-vibration spectra and the electromagnetic
transition rates in $^{13}$C show strong evidence for the occurrence of 
$D_{3h}^{\prime }$ symmetry. The final picture that emerges from our
analysis is that the nucleus $^{13}$C can be considered as a system of three 
$\alpha$-particles in a triangular configuration plus an additional neutron 
moving in the deformed field generated by the cluster, as schematically shown 
in Fig.~\ref{3alphan}. 

Details of our study of $D_{3h}^{\prime }$ as well as results for $T_{d}^{\prime}$ 
will be reported in future publications.

Finally, an important question is the extent to which the cluster structure of
$^{13}$C emerges from microscopic calculations. This nucleus has been 
extensively investigated in the shell model \cite{millener1,lee,millener2} 
where, however, the cluster features are obtained by adjusting the single-particle 
energies, the p-h interactions and the effective charges. In recent years, Fermion 
Molecular Dynamics (FMD) \cite{feldmeier1,roth,neff,feldmeier2} and Antisymmetric 
Molecular Dynamics (AMD) \cite{kanada1,kanada2} have provided 
very detailed and accurate descriptions of light nuclei which confirm the cluster 
structure of $^{12}$C and $^{13}$C obtained from $D_{3h}$ and $D'_{3h}$ symmetry 
(see for example, Fig.~10 of ~\cite{feldmeier2}). Very detailed calculations 
have also been done within the framework of the full four-body $3\alpha+n$ model 
\cite{Yamada} (this reference includes a complete list of microscopic calculations 
of $^{13}$C). It would be of great interest 
to understand whether the cluster structure of $^{12}$C and $^{13}$C emerges from 
{\it ab initio} calculations, such as the no-core shell-model methods (NCCI) 
\cite{navratil1,navratil2,maris} 
for which calculations are planned. The results presented here, based on purely 
symmetry concepts, provide benchmarks for microscopic studies of cluster structure 
of light nuclei.

We thank Dr. Valeria Della Rocca for the preparation of Fig.~\ref{3alphan}. 
This work was supported in part by US Department of Energy Grant
DE-FG-02-91ER-40408 and, in part, by PAPIIT-DGAPA, UNAM Grant IN 109017.

\end{document}